\documentclass{article}
\usepackage{dahlin}

\newcommand\fullI{\mathcal{I}}
\newcommand\fully{\mathbf{y}}

\newcommand\obsy{\mathbf{y}_{\text{o}}}
\newcommand\missy{\mathbf{y}_{\text{m}}}

\newcommand\missyhatk{\widehat{\mathbf{y}}_{k,\text{m}}}
\newcommand\obsI{\mathcal{I}_{\text{o}}}
\newcommand\missI{\mathcal{I}_{\text{m}}}

\title{Inference in Gaussian models with missing data \\ using Equalisation Maximisation}
\author{Johan Dahlin, Fredrik Lindsten and Thomas B. Sch\"{o}n%
\thanks{This work was supported by: the project Calibrating Nonlinear Dynamical Models (Contract number: 621-2010-5876) funded by the Swedish Research Council and CADICS, a Linneaus Center also funded by the Swedish Research Council. The authors are with the Division of Automatic Control, Link{\"o}ping University, Link{\"o}ping, Sweden. E-mail: \{ \texttt{johan.dahlin,lindsten,schon} \}@isy.liu.se.}%
}

\begin{document}
\maketitle
\thispagestyle{empty}
\pagestyle{empty}

\begin{abstract}
Equalisation Maximisation (EqM) is an algorithm for estimating parameters in auto-regressive (AR) models where some fraction of the data is missing. It has previously been shown that the EqM algorithm is a competitive alternative to expectation maximisation, estimating models with equal predictive capability at a lower computational cost. 

The EqM algorithm has previously been motivated as a heuristic. In this paper, we instead show that EqM can be viewed as an approximation of a proximal point algorithm. We also derive the method for the entire class of Gaussian models and exemplify its use for estimation of ARMA models with missing data. The resulting method is evaluated in numerical simulations, resulting in similar results as for the AR processes.
\end{abstract}

\section{Introduction}
A problem in practical applications is that some fraction of the data is missing or cannot be directly observed. This can be the result of faulty sensors, incomplete data records or irregular sampling. It can also be the case that the variables of interest are not directly observable, e.g.\ the latent state process in state space models.

Maximum likelihood (ML) estimation can be readily applied for estimating parameters in many Gaussian models when no data is missing. With the likelihood $\mathcal{L}_{\theta}(\obsy)$ of the observed data $\fully$, the maximum likelihood estimator (MLE) $\widehat{\theta}_{\text{MLE}}$ for some model parametrised by $\theta$ is
\begin{align}
	\widehat{\theta}_{\text{MLE}}  = \arg \max_{\theta \in \Theta} \mathcal{L}_{\theta}(\obsy).
	\label{eq:MLproblem}
\end{align}

To be more specific, let $\fully$ denote a set of data, which is partitioned into \textit{observed data} denoted by $\obsy$ and \textit{missing data} denoted by $\missy$. That is, $\missy = \{y_i: i \in \missI\}$ where $\missI$ denotes the (known) index set of missing data and analogously $\obsy$ is defined using the known index set $\obsI$. As $\{\obsy,\missy\}$ is a partition of $\fully$, we have $\missI \cup \obsI = \fullI = \{1, \ldots, N\}$ where $\fullI$ denotes the full index set.

The observed data is typically conditionally dependent on the missing data. This is the case for many often-used signal processes, such like the auto-regressive (AR) and auto-regressive moving-average (ARMA) processes. Hence, the observed data likelihood $\mathcal{L}_{\theta}(\obsy)$ can be obtained through marginalisation
\begin{align}
	\mathcal{L}_{\theta}(\obsy) &= \dint p_{\theta}(\obsy|\missy) p_{\theta}(\missy) \dd \missy \nonumber \\
	&= \dint p_{\theta}(y_1) \prod_{i=2}^N p_{\theta}(y_i|y_{i-1},\ldots,y_1) \dd \missy.
	\label{eq:latentMLE}
\end{align}

Unfortunately computing, the expression in \eqref{eq:latentMLE} for the use in \eqref{eq:MLproblem} is in many cases difficult or even intractable. This is the result of the high-dimensional integrals involved. One alternative approach is instead to solve the problem in \eqref{eq:latentMLE} is to instead formulate and solve a series of simpler problems, converging to the ML solution. An interesting class of optimisation algorithms that can be used for this purpose is the Proximal Point Algorithms (PPAs). This class of methods is presented in \cite{Rockafellar1976} and \cite{Martinet1970} with discussions in \cite{Chretien2000} for the specific use in MLE.

It is shown in \cite{Chretien2000}, that PPAs targeting the problem \eqref{eq:MLproblem} converges for a wide range of design choices. Specifically, the authors show that the well-known Expectation Maximisation (EM) algorithm, presented in \cite{DempsterLairdRubin1977} and thoroughly discussed in \cite{Mclachlan2007}, is a special instance from the class of PPAs. 

In this paper, we consider another algorithm called Equalisation Maximisation (EqM), introduced in \cite{StoicaXuLi2005a} and \cite{StoicaXuLi2005b}, for iteratively solving the problem in \eqref{eq:MLproblem}. This algorithm is in fact not a member of the class of PPAs, but as we show in the following it is closely related to this class of algorithms and it can be viewed as an approximation of a PPA.

We also generalise the EqM algorithm for estimation of AR model to general Gaussian models and derive the required expressions. This generalised version of the EqM algorithm is exemplified using ARMA models. The resulting implementation using the EqM algorithm is straightforward and simpler than the EM algorithm counterpart.

Finally, we present numerical illustrations from simulation studies in which we infer parameters in AR and ARMA models. These indicate the EqM algorithm can estimate models of comparable predictive ability as the EM algorithm for the ML problem with missing data. Also, we confirm the result presented in \cite{StoicaXuLi2005b}, that the EqM algorithm converges faster than the EM algorithm.

\section{Proximal point algorithms}
The family of PPAs \cite{Rockafellar1976,Martinet1970} consists of algorithms that solves a sequence of regularised optimisation problems. As previously stated, we make use of the algorithm for iteratively solving the ML problem in \eqref{eq:MLproblem}. In this paper, we focus on two choices of the regularisation that iteratively solves \eqref{eq:MLproblem} by creating smaller simpler sub problems.

\subsection{General properties of PPAs}
A PPA is formulated in \cite{Chretien2000} as an iterative maximization procedure, with the following form
\begin{align}
	\widehat{\theta}_{k+1} = \arg \max_{\theta \in \Theta} \left[ \ell_{\theta}(\obsy) - \beta_k d(\theta,	\widehat{\theta}_k) \right],
	\label{eq:PPAoptimisation}
\end{align}
where we have introduced the \textit{observed data log-likelihood} defined as $\ell_{\theta}(\obsy)=\log \mathcal{L}_{\theta}(\obsy)$ and where $\beta_k$ denotes a sequence of (possibly decreasing) positive \textit{relaxation parameters}. Here, $d(\theta,\theta_k)$ denotes a \textit{penalty function}, satisfying $d(\theta,\widehat{\theta}_k) \geq 0$ and $d(\theta,\widehat{\theta}_k)=0$ only if $\theta=\widehat{\theta}_k$.

PPAs have two interesting properties: (i) they monotonically increases the observed data likelihood between two iterations and (ii) they converge to a stationary point of the observed data log-likelihood function. Property (i) follows from the fact that
\begin{align*}
	\ell_{\widehat{\theta}_{k+1}}(\obsy) - \beta_k d(\widehat{\theta}_{k+1},\widehat{\theta}_{k}) 
	\geq
	\ell_{\widehat{\theta}_{k}}(\obsy) - \beta_k d(\widehat{\theta}_{k},\widehat{\theta}_{k}).
\end{align*}
Since $d(\widehat{\theta}_{k},\widehat{\theta}_{k})=0$ and $d(\widehat{\theta}_{k+1},\widehat{\theta}_{k})\geq 0 $, we have
\begin{align*}
	\ell_{\widehat{\theta}_{k+1}}(\obsy) - \ell_{\widehat{\theta}_{k}}(\obsy) 
	\geq 0.
\end{align*}	

For property (ii), assume that the observed data likelihood, the penalty function and their first two derivatives exist and are smooth. In addition assume that $\ell_{\theta}(\obsy)$ is a concave function. Then as the PPA monotonically increases the observed log-likelihood, there exists some $\theta^{\star}$ such that $\widehat{\theta}_{k} \rightarrow \theta^{\star}$ for $k \rightarrow \infty$. This implies that
\begin{align*}
	\theta^{\star} = \arg \max_{\theta \in \Theta} \left[ \ell_{\theta}(\obsy) - \beta_k d(\theta,	\theta^{\star}) \right],
\end{align*}
which means that we have reached a stationary point with respect to $\theta$ and that
\begin{align*}
	\nabla_{\theta} \left[ \ell_{\theta}(\obsy) - \beta_k d(\theta,	\theta^{\star}) \right]_{\theta=\theta^{\star}} =0,
\end{align*}
and as $\nabla_{\theta} [d(\theta,\theta^{\star})]_{\theta=\theta^{\star}}=0$, we have that
\begin{align*}
	\nabla_{\theta} \left[ \ell_{\theta}(\obsy) \right]_{\theta=\theta^{\star}}= 0,
\end{align*}
hence we have reached a stationary point in the observed data log-likelihood function. 

Furthermore, a stronger result can be formulated regarding the convergence rate of the PPA, as presented in \cite{Chretien2000}. Assume that argument in \eqref{eq:PPAoptimisation} fulfils the same assumptions as before and is strictly concave. Also, chose a sequence of positive relaxation parameters converging to zero, when the number of iterations $k$ increases. Then, the problem in \eqref{eq:PPAoptimisation} converges super linearly to the MLE. This result shows that it is possible to create fast algorithms in the class of PPAs for iteratively solving the problem in \eqref{eq:MLproblem}.

\subsection{Expectation maximisation as a PPA}
To apply the PPA for the problem in \eqref{eq:MLproblem}, we need to chose a penalty function so that the sub problems in \eqref{eq:PPAoptimisation} can be easily solved. A possible choice of the penalty function, considered in \cite{Chretien2000}, is the Kullback-Leibler distance between $p_{\theta}(\missy|\obsy)$ and $p_{\widehat{\theta}_{k}}(\missy|\obsy)$, where $\widehat{\theta}_k$ denotes the estimated parameters at iteration $k$. In this case, the penalty is of the following form
\begin{align}
	d_{\text{KL}}(\theta,\widehat{\theta}_{k}) &= \mathbb{E}_{\widehat{\theta}_{k}} \left[ \log \frac{p_{\widehat{\theta}_{k}}(\missy|\obsy)}{p_{\theta}(\missy|\obsy)} \Big| \obsy \right].
	\label{eq:EMpenalty}
\end{align}

By a direct application of Jensen's inequality, we obtain that $d_{\text{KL}}(\theta,\widehat{\theta}_{k}) \geq 0$. We also have the property that $d_{\text{KL}}(\theta,\widehat{\theta}_{k}) = 0$ if and only if $p_{\theta}(\missy|\obsy) \equiv p_{\widehat{\theta}_{k}}(\missy|\obsy)$, i.e.\ $\theta=\widehat{\theta}_{k}$. Hence, this is a valid choice for the penalty function in a PPA. With the choice $\beta_k \equiv 1$ for every $k$, we get
\begin{align*}
	\widehat{\theta}_{k+1} 
	&= \arg \max_{\theta \in \Theta} \left\{ \ell_{\theta}(\obsy) + \mathbb{E}_{\widehat{\theta}_{k}} \left[ \log \frac{ p_{\theta}(\missy|\obsy)}{p_{\widehat{\theta}_{k}}(\missy|\obsy)} \Big| \obsy \right] \right\} \nonumber \\
	&=  \arg \max_{\theta \in \Theta} \left\{ \ell_{\theta}(\obsy) + \mathbb{E}_{\widehat{\theta}_{k}} \left[\log p_{\theta}(\missy|\obsy) \big| \obsy \right] \right\} \nonumber \\
	&= \arg \max_{\theta \in \Theta} \mathcal{Q}(\theta,\widehat{\theta}_{k}),
\end{align*}
where the second equality is a result of the fact that $\log p_{\widehat{\theta}_{k}}(\missy|\obsy)$ is independent of $\theta$ and can be neglected in the optimisation. The function $\mathcal{Q}(\theta,\widehat{\theta}_{k})$ is commonly referred to as the \textit{intermediate quantity} in the EM algorithm. This quantity is defined by
\begin{align*}
	\mathcal{Q}(\theta,\widehat{\theta}_{k})
	&= \mathbb{E}_{\widehat{\theta}_{k}} \left[\log p_{\theta}(\missy,\obsy) \big| \obsy \right] \nonumber \\
	&= \ell_{\theta}(\obsy) + \mathbb{E}_{\widehat{\theta}_{k}} \left[\log p_{\theta}(\missy|\obsy) \big| \obsy \right],
\end{align*}
which follows from taking the conditional expectation with respect to $p_{\widehat{\theta}_k}(\missy|\obsy)$ of the following factorisation of the log-likelihood of the observed data
\begin{align}
	\ell_{\theta}(\obsy) = 	\log p_{\theta}(\missy,\obsy) - \log p_{\theta}(\missy|\obsy),
	\label{eq:MLEmissingDecomp}
\end{align}
which is equivalent to the expression in \eqref{eq:latentMLE}.

Hence, this specific choice of the penalty function and relaxation parameters results in the well-known EM algorithm. We recover the well-known results regarding the EM algorithm, by the previous discussion and its PPA formulation. The first property is that the EM algorithm is monotonically increasing the observed data likelihood in each iteration. The second property is that the EM algorithm converges to the stationary point of the observed data log-likelihood, denoted $\theta^{\star}=\widehat{\theta}_{\text{EM}}$.

Also, as the relaxation parameter is fixed and does not decrease with time, the convergence rate can be shown to be linear. An interesting problem is therefore to reformulate the EM algorithm using a PPA with super linear convergence rate. For more discussion on this problem and some results, see \cite{Chretien2000}.

\section{Equalisation maximisation}
For many problems, the expectation operator in \eqref{eq:EMpenalty} is computationally costly to evaluate. Therefore it can in many cases be beneficial to consider alternatives of this regularisation to decrease the computational cost or to increase the convergence rate. Especially, as it is well-known that the EM algorithm has a slow overall convergence rate.

In this section, we consider another, simpler, choice of regularisation. The resulting method is the EqM algorithm, first derived in \cite{StoicaXuLi2005a} and \cite{StoicaXuLi2005b}, a type of cyclic maximisation algorithm. In this second, we sow how to apply the EqM algorithm for the class of Gaussian processes. In particular, we derive an algorithm for parameter estimation in ARMA processes with missing data.

\subsection{Connection to PPA and general properties}
If the missing data were available then the optimisation of $\log p_{\theta}(\missy,\obsy)$ in \eqref{eq:MLEmissingDecomp} is rather easy. This corresponds to the case of fully observed data and can be solved using standard ML methods, e.g.\ using numerical optimisation of the likelihood function with the Gauss-Newton algorithm. Hence, this is a much simpler problem to solve than \eqref{eq:MLproblem} with the observed data log-likelihood in \eqref{eq:latentMLE}.

Suppose that we have an estimator of the missing data $\missyhatk=\missyhatk(\widehat{\theta}_{k-1},\obsy)$, given the current estimate of the parameters and the observed data. Inserting this into \eqref{eq:MLEmissingDecomp} gives
\begin{align}
	 \log p_{\theta}(\missyhatk,\obsy) = \ell_{\theta}(\obsy) + \log p_{\theta}(\missyhatk|\obsy),
	 \label{eq:EqMrelation}
\end{align}
and from \eqref{eq:MLproblem} we obtain an iterative algorithm as
\begin{align}
	\widehat{\theta}_{k+1} = \arg \max_{\theta \in \Theta} \left[ \ell_{\theta}(\obsy) + \log p_{\theta}(\missyhatk|\obsy) \right],
	\label{eq:EqMPPA}
\end{align}
which is the EqM algorithm presented in \cite{StoicaXuLi2005a} and \cite{StoicaXuLi2005b}. The expression resembles the PPA in \eqref{eq:PPAoptimisation} and is obtained by letting $\beta_k \equiv 1$ and using the following function
\begin{align}
	d_{\text{EqM}}(\theta,\widehat{\theta}_k) = - \log p_{\theta}(\missyhatk|\obsy) + c_{\text{EqM}},
	\label{eq:EqMpenalty}
\end{align}
for some constant $c_{\text{EqM}}$. For \eqref{eq:EqMPPA} to be a PPA we require that $d_{\text{EqM}}(\theta,\widehat{\theta}_k)$ is a penalty function. Therefore we would need that $d_{\text{EqM}}(\theta,\theta)=0$ and that $d_{\text{EqM}}(\theta,\theta)$ is independent of $\theta$. This can be fulfilled by choosing $c_{\text{EqM}}$ as
\begin{align}
	c_{\text{EqM}} = \log p_{\theta}(\missyhatk(\theta,\obsy)|\obsy),
	\label{eq:EqMc}
\end{align}
which in the following is shown to be independent of $\theta$. The expression in \eqref{eq:EqMpenalty} with this choice of constant is referred to as the \textit{equalisation function} in \cite{StoicaXuLi2005b}, hence the name of the algorithm.

Unfortunately, this equalisation function does not necessarily fulfil $d_{\text{EqM}}(\theta,\widehat{\theta}_k) \leq 0$. Therefore this choice of relaxation parameters and penalty function does not fulfil the requirements for \eqref{eq:EqMpenalty} to be a PPA. Hence, the monotonicity and convergence properties previously discuss does not follow for the EqM algorithm. Later, we however show by numerical simulations that this choice results in a convergent algorithm with a behaviour similar to the EM algorithm.

It is shown in \cite{StoicaXuLi2005b} that the EqM algorithm converges to another point than the MLE. This can be easily seen using the PPA formulation and the corresponding convergence discussion. Under the same assumptions as for the PPA and assuming that the EqM algorithm converges to some parameters $\theta_{\text{EqM}}$ when $k \rightarrow \infty$, we have
\begin{align*}
	\frac{\partial \ell_{\theta}(\obsy)}{\partial \theta} \Big|_{\theta=\theta_{\text{EqM}}}
	+ \frac{\partial \log p_{\theta}(\missyhatk|\obsy)}{\partial \theta} 
	\Big|_{\begin{subarray}{l} 
	\missyhatk=\missyhatk(\widehat{\theta}_{\text{EqM}},\obsy) \\ \theta=\theta_{\text{EqM}} 
	\end{subarray}}=0,
\end{align*}
which shows that the EqM algorithm does not necessarily converges to the same parameters as the EM algorithm, i.e.\ $\theta_{\text{EqM}} \neq \theta_{\text{EM}}$. Note that, the difference between the parameter estimates depends on the derivative of the equalisation function with respect to $\theta$.

\subsection{Parameter inference in Gaussian models}
To make use of the EqM algorithm, we need to find an expression for the equalisation function. In this paper, we limit ourselves to parameter inference in Gaussian models. It is however, possible to find similar expressions for other models than the Gaussian. 

A stochastic process $\fully = \{y_t: t \in \fullI\}$ is a \textit{Gaussian process}, if for any subset $\mathbf{y}_s$ of $\fully$, the resulting random vector is distributed as a multivariate Gaussian distribution
\begin{align*}
	\mathbf{y}_s \sim \mathcal{N}(\mathbf{y}_s; \mu, \Sigma),
\end{align*}
with mean $\mu=\mathbb{E}[\mathbf{y}_s]$ and covariance matrix $\Sigma=\mathsf{Cov}[\mathbf{y}_s,\mathbf{y}_s]$. In the following, the subsets of $\fully$ are the previously defined partition of the data into $\{\obsy,\missy\}$.

The assumption that the process is Gaussian implies that $\obsy$ and $\missy$ are jointly Gaussian. It then follows that the conditional distribution of $\missy$ given $\obsy$ is Gaussian with conditional mean $\mu_{\theta}(\missy|\obsy)$ and conditional covariance matrix $\Sigma_{\theta}(\missy|\obsy)$. Guided by the structure of the problem and the relation in \eqref{eq:EqMpenalty}, we chose the following equalisation function
\begin{align}
	\missyhatk &= \mu(\missy|\obsy) \nonumber \\
	&+ \Sigma_{1,\cdot}(\missy|\obsy) \left[ \frac{ \log |\widehat{\theta}_0| - \log |\Sigma(\missy|\obsy)|}{\Sigma_{1,1}(\missy|\obsy)}\right]^{1/2},
	\label{eq:EqFuncGaussian}
\end{align}
which is discussed in \cite{StoicaXuLi2005b} and where we have suppressed the dependence of $\theta$ for brevity. Here, $\Sigma_{1,\cdot}(\missy|\obsy)$ and $\Sigma_{1,1}(\missy|\obsy)$ denote the first row and the first element of the conditional covariance matrix, respectively. Finally, $\widehat{\theta}_0$ denotes a user-defined constant which according to \cite{StoicaXuLi2005b} can be chosen to unity without any loss of generality.

Inserting this choice of equalisation function into the conditional distribution in 
\eqref{eq:EqMrelation} gives
\begin{align*}
		\log p_{\theta}(\missyhatk|\obsy) = \left(2 \pi \right)^{-|\missI|/2} |\widehat{\theta}_0|^{-1/2}, = c_{\text{EqM}},
\end{align*}
which is independent of $\theta$ as required.

\subsection{Parameter inference in ARMA models}
In this section, we exemplify the EqM algorithm by applying it to ARMA processes. A process of this type with orders ($p,q$) has the structure
\begin{align}
	y_t + \sum_{k=1}^p \phi_k y_{t-k} = e_t + \sum_{k=1}^q \lambda_k e_{t-k},
	\label{eq:ARMAprocess}
\end{align}
where $e_t \sim \mathcal{N}(0,\sigma_e^2)$ and $y_t$ denotes the process value at time $t$. Here, $\phi_k$ and $\lambda_k$ are the model parameters, i.e.\ we assume in the following that the model orders ($p,q$) are known. Note that this ARMA process has the AR process of order $p$ and the moving-average (MA) process of order $q$, respectively as special cases. These cases are recovered if $\lambda_k=0$ or $\phi_k=0$ for all $k$, respectively. 

To apply the EqM algorithm, we need to rewrite \eqref{eq:ARMAprocess} on a matrix form. This can be done by defining the following Gaussian process
\begin{align}
	\fully \sim \mathcal{N}(\fully;0,\Psi \Psi^{\top} \sigma_e^2),
	\label{eq:ARMAmatrix}
\end{align}
where $\Psi^{\top} = \Lambda \Phi^{-1}$ and the $N \times N$-matrix of AR parameters $\Phi$ is defined by
\begin{align}
\Phi^{-\top} &= \begin{bmatrix}
	1 & \phi_1 & \phi_2 & \ldots & 0 & 0 \\
	0 & 1 & \phi_1 & \ldots & 0 & 0 \\
	0 & 0 & 1 & \ldots & 0 & 0 \\
	\vdots & \vdots & \vdots & \ddots & \phi_r & 0 \\
	0 & 0 & 0 & \ldots & \phi_{r-1} & \phi_r		
	\end{bmatrix}. \label{eq:EqMphiMatrix}
\end{align}
The second $N \times N$-matrix of MA parameters $\Lambda$ is defined analogously to $\Phi$ by replacing $\phi_k$ with $\lambda_k$. 

An equivalent form of \eqref{eq:ARMAprocess} which we also make use of is the state space representation with the structure
\begin{subequations}
\begin{align}
	x_t &= 
	\underbrace{\begin{bmatrix}
		0 & 1 & 0 & \ldots & 0 \\
		0 & 0 & 1 & \ldots & 0 \\
		\vdots & \vdots & \vdots & \ddots & \vdots \\
		0 & 0 & 0 & \ldots & 1 \\
		-\phi_r & -\phi_{r-1} & -\phi_{r-2} & \ldots & -\phi_1 
	\end{bmatrix}}_{\triangleq A(\theta)}
	x_{t-1} + 
	v_t, \\
	y_t &= 
	\begin{bmatrix}
		1 & 0 & \ldots & 0 	
	\end{bmatrix}
	x_t	+ e_t,
\end{align}%
\label{eq:ARMAssmJones}%
\noindent with $r=\max(p,q+1)$ and where $y_t$ and $x_t$ denotes the system output and state at time $t$, respectively. Also, the two independent noise processes are denoted $v_t \sim \mathcal{N}(0,Q(\theta))$ and $e_t \sim \mathcal{N}(0,R(\theta))$, with $Q(\theta)=B(\theta)B(\theta)^{\top}\sigma^2_v$ and $R(\theta)=\sigma^2_e$ using $B(\theta)=[1, g_1, \ldots, g_r]^{\top}$. The coefficients $g_k$ are obtained by the following recursion
\end{subequations}%
\begin{align*}
	g_k = \lambda_{k-1} - \sum_{j=1}^{k-1} \phi_j g_{k-j},
\end{align*}
using $g_1=1$ and $\lambda_k = 0$ for $k > q$. 

The complete EqM algorithm for estimating the parameters in an ARMA($p,q$) model defined by \eqref{eq:ARMAmatrix} is given in Algorithm~\ref{alg:EqMARMA}. 

The EqM algorithm for the estimation of an AR($p$) model is obtained by using $\Theta=\mathbf{I}_N$, i.e.\ an $N \times N$ identity matrix, in \eqref{eq:ARMAmatrix}. For this case the expressions for the conditional mean and conditional covariance matrix in Algorithm~\ref{alg:EqMARMA} remains the same. If $\Phi=\mathbf{I}_N$ in \eqref{eq:ARMAmatrix}, we instead obtain an algorithm for estimating MA($q$) models. In this case, the expressions for the conditional mean and conditional covariance matrix are more easily obtained by standard Gaussian conditioning using the information matrix form.

\begin{algorithm}[!t]
\caption{EqM for ARMA($p,q$) with missing data.}
\begin{small}
Assume that $\widehat{\theta}_0$ is available together with some observed data $\obsy$. Set $k=0$. Repeat the following until convergence:
	\begin{enumerate}
		\item Construct the $\Phi$ and $\Lambda$ matrices as described in \eqref{eq:EqMphiMatrix} using $\widehat{\theta}_k$.
		\item Calculate $\Sigma = \Phi \Lambda^{-1}$ and partition $\Sigma$ into 
		\begin{align*}
			\Sigma = \begin{bmatrix} \Sigma_{\text{mm}} & \Sigma_{\text{mo}} \\ \Sigma_{\text{om}} & \Sigma_{\text{oo}} \end{bmatrix},
		\end{align*}
		where $\Sigma_{\text{mm}}$ and $\Sigma_{\text{oo}}$ denote the covariance matrix of the missing data, $\missy$, and observed data, $\obsy$, respectively.
		\item Compute the conditional mean and covariance using
		\begin{align*}
			\mu_{\theta}(\missy|\obsy)
			&= \Sigma_{\text{mo}} \Sigma_{\text{oo}}^{-1} \mathbf{y}, \\
			\Sigma_{\theta}(\missy|\obsy)
			&= \Sigma_{\text{mm}} - \Sigma_{\text{mo}} \Sigma_{\text{oo}}^{-1} \Sigma_{\text{om}}.
		\end{align*}
		\item Estimate the missing data with \eqref{eq:EqFuncGaussian} and use this to estimate the new parameters $\widehat{\theta}_{k+1}$ using
		\begin{align*}
			\widehat{\theta}_{k+1} = \arg \max_{\theta \in \Theta} p_{\theta}(\missyhatk,\obsy),
		\end{align*}
		with $\missyhatk=\missyhatk(\widehat{\theta}_{k},\obsy)$.
		\item Set $k=k+1$.
	\end{enumerate}
\end{small}
\label{alg:EqMARMA}
\end{algorithm}

\section{Numerical illustrations}
In this section, we provide three different numerical illustrations of using EqM as presented in Algorithm \ref{alg:EqMARMA}, for parameter inference in Gaussian processes when some fraction of the data is missing. 

\subsection{Experimental set-up}
In the first two illustrations, we are generating a large number of stable random AR($p$) and ARMA($2,2$) processes, where $p \in \{1,\ldots,15\}$. From each random process, we sample a single realisation of $N$ time steps and partition it into an estimation set using $2/3$ of the data and a validation set using the remaining $1/3$ of the data. From the estimation set, a random fraction of the data is removed, such that a certain fraction of the data is missing. The remaining observed estimation data is used to estimate the model parameters (assuming known model orders) and construct one-step-ahead predictors. These predictors are used to compute the model fit on the validation data.

In all cases, we compare the results from the EqM algorithm with the solution obtained using the EM algorithm. For the AR models, we use the same algorithm as in \cite{StoicaXuLi2005b}. For the ARMA models, we make use of the algorithm from \cite{ShumwayStoffer1982} which is thoroughly discussed in \cite{shumway2010time}. This is not an explicit EM algorithm for ARMA processes but to the authors' knowledge, no such method currently exists in the literature. We instead estimate a parametrised state space model, in which $A(\theta)$, $Q(\theta)$ and $R(\theta)$ from \eqref{eq:ARMAssmJones} are fully parametrised.

Both the EqM and EM algorithms are initialised using a \textit{naive} estimator, in which the models are estimated assuming that all the missing data is zero. Also, in all simulations we use a known initial state, $y_i=0$ for $i=0,-1,\ldots$

\subsection{AR($p$) simulations}
We generate one realisation of $N=1\thinspace 250$ data points from each of the $250$ randomly generated stable AR($p$) processes with $\sigma_e^2=1$. Both algorithms are executed for a maximum of $100$ iterations, using the convergence criteria $\|\widehat{\theta}_k-\widehat{\theta}_{k-1}\|_2 \leq 10^{-6}$, i.e.\ a small difference in the estimated parameter vectors between two consecutive iterations.

\begin{figure}[!t]
	\centering
	\includegraphics[width=0.9\textwidth]{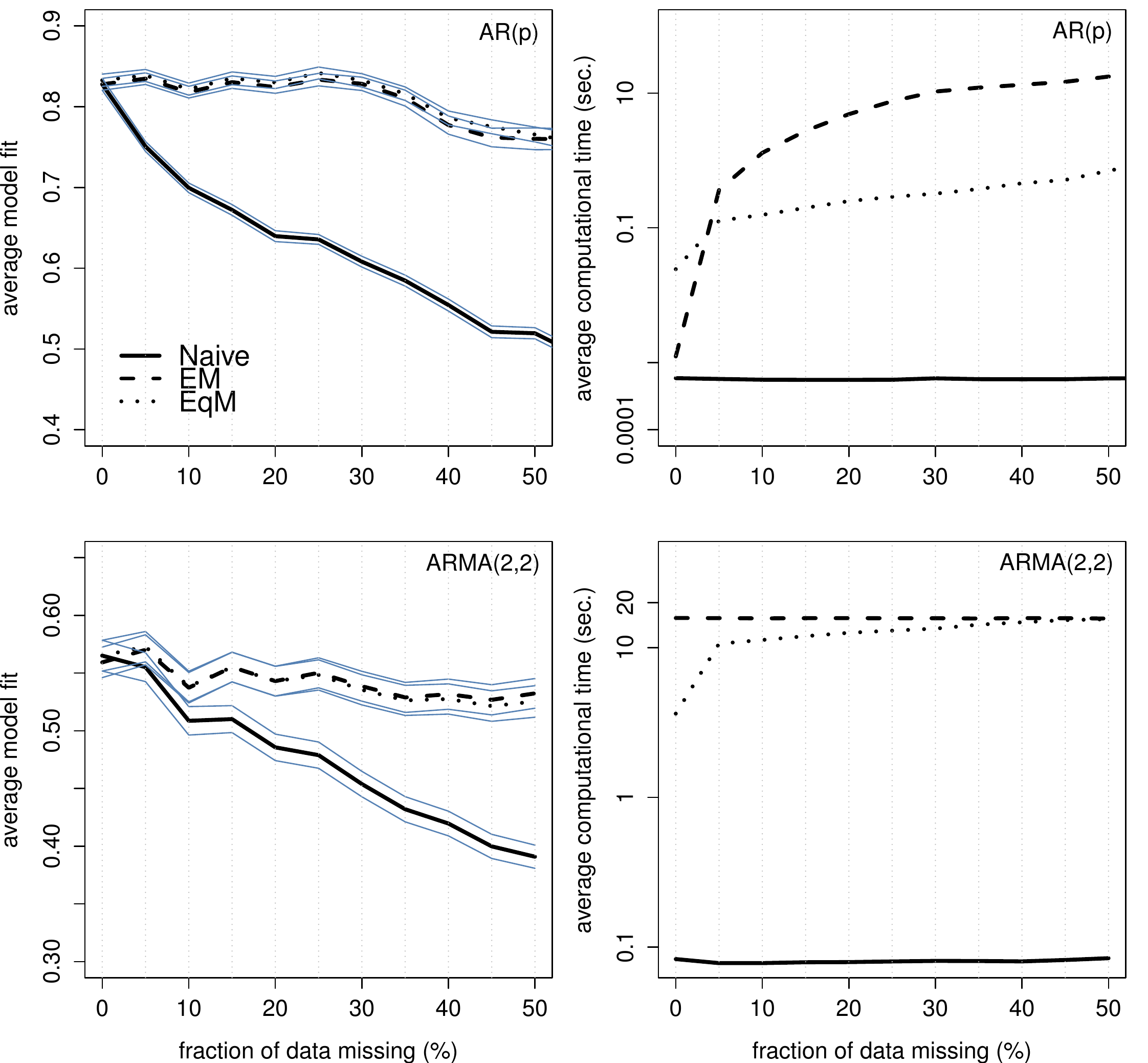}
	\caption{Upper left: the average model fit from $250$ random AR($p$) processes with the order $p$ randomly selected from $\{1,2,\ldots,15\}$. The blue lines indicate the $95 \%$ confidence intervals. Upper right: the average computational time until convergence. Lower: the average model fit and computational time for $2 \thinspace 000$ random ARMA(2,2) processes.}
	\label{fig:simStudies}
\end{figure}

In the upper part of Figure~\ref{fig:simStudies}, the average model fit and the average computational time per estimated model are reported for different fractions of missing data. The average model fit is comparable for both algorithms with no statistical significant difference. For the naive method, the model fit decreases rather quickly with increasing fraction of missing data. At the same time, the model fit remains almost constant for the other two approaches. 

We note that the main difference between EqM and EM lies in the average computational time. The EqM algorithm is often at least one order of magnitude faster than the EM algorithm. The computational time for both methods seem to scale similarly as the fraction of missing data increases.

\subsection{ARMA(2,2) simulations}
To study the behaviour of the EqM algorithm in estimating ARMA models, we generate $2\thinspace 000$ random stable ARMA(2,2) processes. From each process we sampled one realisation of $N=1\thinspace500$ data points using the noise variance $\sigma^2_e=0.01$. Both the EM and EqM algorithms are in this case executed for exactly $50$ iterations, i.e.\ without any other convergence criteria.

The result is presented in the lower part of Figure~\ref{fig:simStudies}, which is quite similar to the case study in the previous section. The model fit is comparable for both methods with no statistical significant difference. As before, the model fit is almost kept constant by the EqM and EM algorithms, but falls of quickly for the naive approach. 

Also the average computational time per system is again larger for the EM algorithm, but only with a factor of about $2$. In difference with the previous case, the computational time used by the EqM algorithm approaches the EM algorithm. This makes the two algorithms comparable at the higher fractions of missing data considered in this illustration.

\subsection{A specific ARMA(2,2) process}
Lastly, we study a specific realisation of an ARMA(2,2) process with the parameters $\{\phi_1,\phi_2,\lambda_1,\lambda_2\} =$ $\{-0.8897,0.4858,-0.2279,0.2488\}$. A realisation of $N=1\thinspace 500$ data points is sampled using $\sigma^2_e=0.1$. Using this specific batch of data, we randomly remove data points in $50$ Monte Carlo runs such that a certain fraction of the estimation data is missing. The EqM and EM algorithms are both executed for a maximum of $50$ iterations with the convergence criteria $\| \ell_{\widehat{\theta}_k}(\obsy) - \ell_{\widehat{\theta}_{k-1}}(\obsy) \|_2 \leq 10^{-6}$, i.e.\ a small difference in the observed data log-likelihood between two consecutive iterations.

\begin{figure}[!t]
	\centering
	\includegraphics[width=0.9\textwidth]{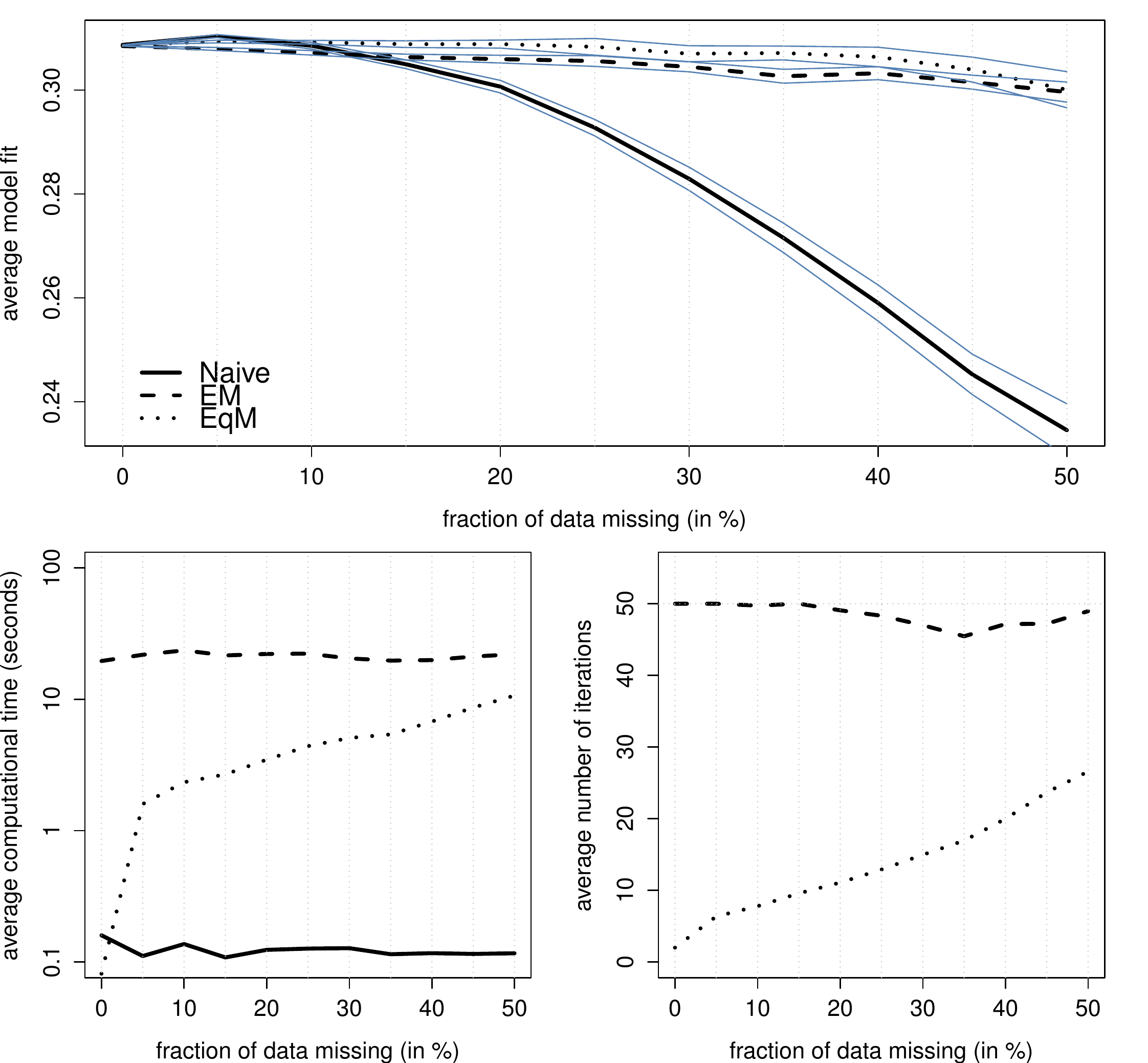}
	\caption{Upper: average model fit with $95\%$ confidence intervals from $50$ Monte Carlo runs for a specific ARMA($2,2$) model. Lower left: the average computational time per estimated model. Lower right: the average number of iterations until the observed log-likelihood changes by less than $10^{-6}$ between two consecutive iterations.}
	\label{fig:ARMA22case}
\end{figure}

The results are presented in Figure~\ref{fig:ARMA22case}. We see that the model fit is again similar for the EqM and EM algorithms. However, the number of iterations until convergence is lower for the EqM algorithm, but increases when larger fractions of data are missing. The same behaviour is visible from the computational time needed for the EqM algorithm to converge. As the fraction of missing data increases, the EqM approaches the computational time required by the EM algorithm. Still, EqM converges in fewer iterations than the EM algorithm, making each iteration of the former more costly than of the latter.

\begin{figure}[!t]
	\centering
	\includegraphics[width=0.9\textwidth]{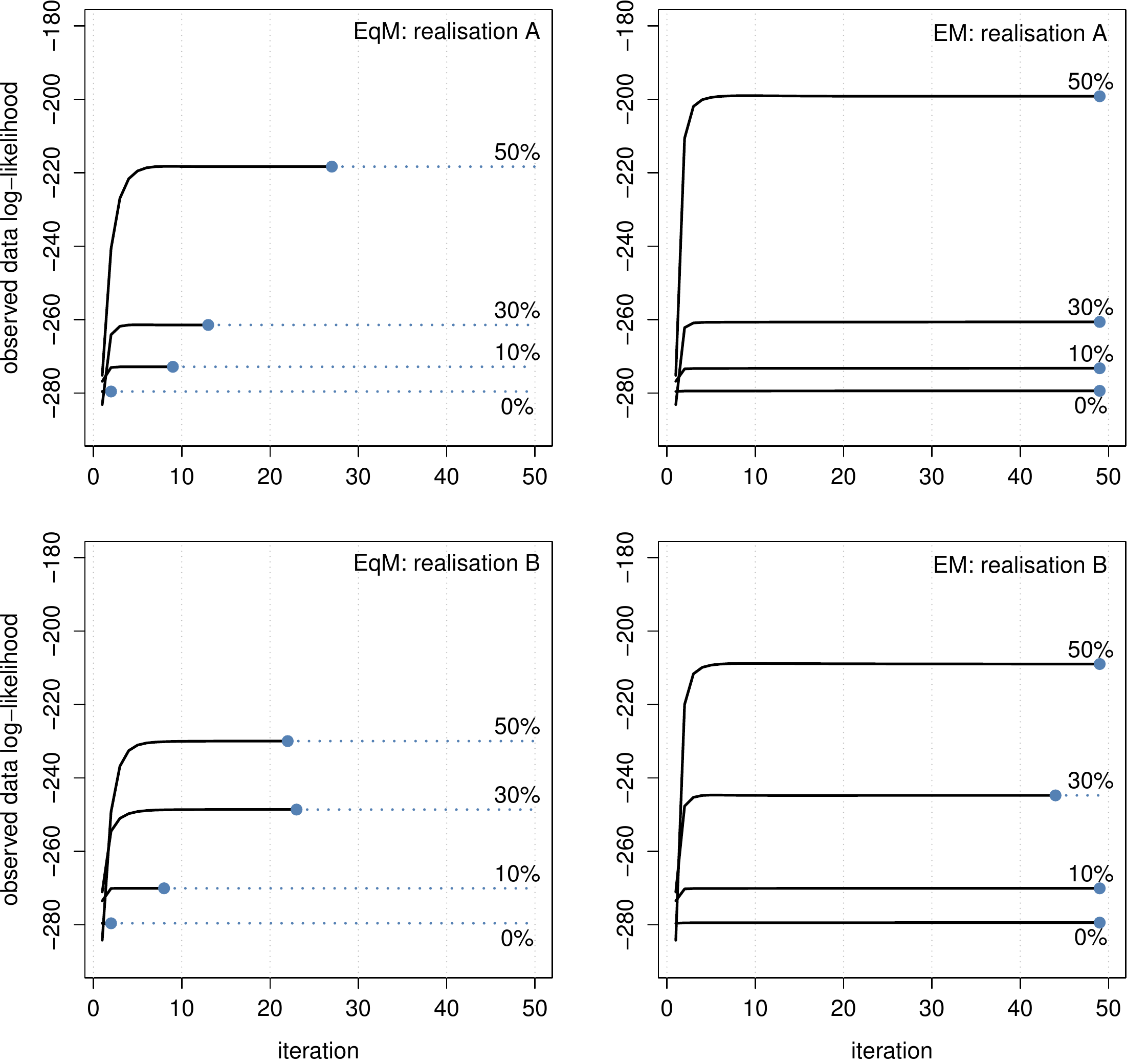}
	\caption{The observed log-likelihood of the estimated ARMA($2,2$) model for the EqM and EM algorithm from two Monte Carlo runs. The text labels in connection with the lines indicate the fraction of data missing from the realisation considered. The circles indicate the iteration at which the parameters converges or when maximum number of iterations is reached.}
	\label{fig:ARMA22caseLL}
\end{figure}

In Figure \ref{fig:ARMA22caseLL}, we study two specific Monte Carlo runs from the experiment and the convergence of the observed data log-likelihood. The iteration at which the algorithm finished is indicated with filled circles, after which the observed data log-likelihood is shown as constant with dashed lines. 

We see that the EqM algorithm often typically increases the log-likelihood monotonically. This cannot be guaranteed by the previous analysis, but it seems that in practice this property often holds true. However, the authors have observed cases when the observed data log-likelihood decreases between two iterations. These changes are often minor and overall the EqM algorithm seems to increase the observed data log-likelihood between iterations.

For smaller fractions of missing data, the EqM and EM algorithms seem to converge to comparable observed log-likelihoods. This indicates that they converge to similar points in the parameter space, in that they have the similar likelihoods. As the previous analysis discussed, there exists a \textit{gap} between the observed data log-likelihoods between the EqM solution and the MLE solution. However, it seems that this gap is small when only small fractions of data are missing but increases when more data is removed. As previously shown, the EqM algorithm still estimates models of equal predictive power as the EM solution.

It is difficult to generalise the results regarding the convergence rate from these specific cases. In \cite{StoicaXuLi2005b}, it is stated that the EqM algorithm converges more rapidly than the EM algorithm. This is also true in this case, i.e.\ the converge criteria is reached faster for the EqM algorithm. The rapid increase in the observed data log-likelihood is comparable between the EqM and EM algorithms for the first few iterations. This indicates that the EqM algorithm enjoy the initial rapid behaviour of the EM algorithm, but also enjoys a faster convergence after the initial phase than the EM algorithm.

\section{Concluding remarks}
We have compared the EqM algorithm with the EM algorithm on randomly generated AR and ARMA processes. Both algorithms are equivalent, comparing the predictive capability of the estimated models, when the fraction of data missing ranges between $0\%$ and $50\%$. 

The major difference is the computational time and the number of iterations required for convergence. The EqM algorithm is on average $10$ and $2$ times faster when estimating AR and ARMA models, respectively. The same result holds for the number of iterations required for convergence.

The observed data log-likelihood is comparable between the EqM and EM algorithm, when the fraction of missing data is less than say $30 \%$. When the amount of missing data increases beyond that, the EqM algorithm estimates models with a lower observed data log-likelihood than the EM algorithm.

Future work includes a more detailed convergence analysis of the EqM algorithm, to better be able to quantify the gap between the parameter estimates and the MLE. Also, it would be interesting to investigate if the EqM algorithm can be applied to linear Gaussian state space models or other classes of processes than the Gaussian. 

\bibliographystyle{IEEEbib}
\bibliography{../eqm-arma-refs}
\end{document}